
\input phyzzx
\nonstopmode
\sequentialequations
\tolerance=5000
\overfullrule=0pt
\nopubblock
\twelvepoint

\line{\hfill PUPT 1570 , IASSNS-HEP 95/76}
\line{\hfill cond-mat/9510132}
\line{\hfill October 1995}
\titlepage
\title{Spin Singlet Ordering Suggested by Repulsive Interactions}

\vskip.2cm
\author{Chetan Nayak\foot{Research supported in part by a Fannie
and John Hertz Foundation fellowship.~~~
nayak@puhep1.princeton.edu}}
\vskip.2cm
\centerline{{\it Department of Physics}}
\centerline{{\it Joseph Henry Laboratories}}
\centerline{{\it Princeton University}}
\centerline{{\it Princeton, N.J. 08544}}

\author{Frank Wilczek\foot{Research supported in part by DOE grant
DE-FG02-90ER40542.~~~wilczek@sns.ias.edu}}

\vskip.2cm
\centerline{{\it School of Natural Sciences}}
\centerline{{\it Institute for Advanced Study}}
\centerline{{\it Olden Lane}}
\centerline{{\it Princeton, N.J. 08540}}
\endpage

\abstract{We consider a correlated wavefunction including particle-hole
pairing at half a reciprocal lattice vector for itinerant electrons hopping
on a square lattice in two dimensions and subject both to
on-site and nearest-neighbor repulsion.  We find, within mean field theory,
that in a suitable range
of parameters there is a tendency toward a novel form of ordering
characterized by discrete symmetry breaking and
a gap at the magnetic zone boundary, which supports  hole-like
semiconductor behavior near half filling despite an apparently normal
Fermi surface.  The effective quasiparticle couplings to phonons includes
a significant d-wave piece, which plausibly leads to an especially robust
tendency toward superconductivity.}

\endpage

\REF\schrieffer{J.R. Schrieffer, X.G. Wen, and S.C. Zhang, Phys. Rev.
{\bf B 39} (1989) 11663}

\REF\brout{For a nice systematic exposition of this point of view
see  R. Brout, {\it Phase Transitions\/}, (W.A. Benjamin, New York 1965)}

\REF\millis{A.J. Millis, Phys. Rev. {\bf B 48} (1993) 7183}

\REF\sachdev{S.Sachdev, A.V. Chubukov, and A. Sokol, {\it Crossover
and scaling in a nearly antiferromagnetic Fermi liquid in two
dimensions},
Yale preprint (1994). }

One has become accustomed, especially in connection with the superfluid
phases of He$^3$, to the existence of quite intricate pairing correlations
between particles.  On the other hand, well-known forms of ordering
including ferromagnetic, antiferromagnetic, and charge-density wave can
be described in terms of particle-hole pairing [\schrieffer,\brout ].
It is quite natural, then,
to consider by analogy the possibility of states of matter characterized by
particle-hole pairing of more general types.   In this note we shall show
that indeed one such ordering is suggested to be favorable
in an analysis of the mean-field
correlation energy induced by very simple, purely repulsive interactions.
This may have significant implications for the Mott insulator problem, as
we shall discuss.

\chapter{Analysis}

Let us consider electrons on a square lattice in two dimensions with
spacing $a$. We suppose that the screened Coulomb interaction can be
modelled by an on-site repulsion of magnitude $U$
and nearest-neighbor repulsion of magnitude $V$, that is:
$$
H_{\rm int.} ~=~
U \sum_x
c^\dagger_\uparrow (x) c_\uparrow (x)\,c^\dagger_\downarrow (x) c_\downarrow
(x)
\,\,+\,\, {1\over 2} V \sum_{x,x^\prime}
c^\dagger_\alpha (x) c_\alpha (x)\,
c^\dagger_\beta (x^\prime ) c_\beta (x^\prime ) ~,
\eqn\interaction
$$
where $x^\prime$ runs over the four nearest neighbors of $x$,
and $\alpha,\beta=\uparrow,\downarrow$ are summed over.
Transforming \interaction\ to momentum space, we find the form
$$\eqalign{
H_{\rm int.} ~=~
&U \sum_{k_i}
c^\dagger_\uparrow (k_1) c_\uparrow (k_2)
c^\dagger_\downarrow (k_3) c_\downarrow (k_4) \,
\delta_L (k_1+k_3-k_2-k_4)\cr
&+ 2V\sum_{k_i}
c^\dagger_\alpha (k_1) c_\alpha (k_2)\,
c^\dagger_\beta (k_3) c_\beta (k_4 )\,
 C({{\bf k}_3}-{{\bf k}_4})\, \delta_L (k_1+k_3-k_2-k_4)~,
\cr}\eqn\kint$$
where $\delta_L$ denotes the delta function in the reciprocal lattice,
and  $C({{\bf k}_3}-{{\bf k}_4})=
\cos (k_3-k_4)_x a ~+~ \cos (k_3 -k_4)_y a$.

Interesting possibilities for particle-hole correlations, whose energetic
consequences can be seen
in a mean-field analysis of \kint , involve ordering at momentum transfer
$Q$, where $2Q$ is a reciprocal lattice vector.  We shall focus on the
case $Q = (\pi/a,\pi/a)$.
In the spirit of BCS theory, we postulate a wave-function of the form
$$
\Psi ~=~ \prod_k \, (u_{k\uparrow} c^\dagger_{k\uparrow}
+ v_{k\uparrow} c^\dagger_{k+Q \uparrow} )\,
(u_{k\downarrow} c^\dagger_{k\downarrow}
+ v_{k\downarrow} c^\dagger_{k+Q \downarrow} )\, |0 > ~,
\eqn\wavefnct
$$
where the product runs over some subset of the interior of the
magnetic zone (i.e. the diamond $(k_x \pm k_y )a = \pm \pi$),
and $|u|^2 + |v|^2 =1$ for all
values of the indices, to preserve normalization, and the number of
factors is determined by the density of electrons (see below).
The expectation value of $H_{\rm int.}$ in the state $\Psi$ is
$$\eqalign{
\langle \Psi | H_{\rm int.} | \Psi \rangle =
 {\sum_{k,k^\prime}}\,\,&U(\bar u_{k\uparrow} v_{k\uparrow} +
u_{k\uparrow} \bar v_{k\uparrow} )
 (\bar u_{k^\prime \downarrow} v_{k^\prime\downarrow}
   + u_{k^\prime\downarrow} \bar v_{k^\prime\downarrow})\cr
&-\, 2V(\bar u_{k\uparrow} v_{k\uparrow} + u_{k\uparrow} \bar v_{k\uparrow}
 +\bar u_{k\downarrow} v_{k\downarrow} + u_{k\downarrow} \bar v_{k\downarrow})
\cr
&\qquad\qquad\times
(\bar u_{k^\prime\uparrow} v_{k^\prime\uparrow}
   + u_{k^\prime\uparrow} \bar v_{k^\prime\uparrow}
 +\bar u_{k^\prime\downarrow} v_{k^\prime\downarrow}
   + u_{k^\prime\downarrow} \bar v_{k^\prime\downarrow} )\cr
&+\, V\,
C({\bf k}-{\bf k^\prime})\,
 \bigl(
  (\bar u_{k\uparrow}v_{k\uparrow} - u_{k\uparrow}\bar v_{k\uparrow} )
  (\bar u_{k^\prime\uparrow}v_{k^\prime\uparrow}
    - u_{k^\prime\uparrow}\bar v_{k^\prime\uparrow} )
\cr
&\qquad\qquad\qquad+
  (\bar u_{k\downarrow}v_{k\downarrow} - u_{k\downarrow}\bar v_{k\downarrow} )
  (\bar u_{k^\prime\downarrow}v_{k^\prime\downarrow}
    - u_{k^\prime\downarrow}\bar v_{k^\prime\downarrow} ) \bigr)~,
\cr}
\eqn\expectv$$
where the bar denotes complex conjugate.

Without loss of generality we
may take the $u$s to be real and non-negative, and write
$v_{k\alpha} \equiv w_{k\alpha} \exp (i\phi_{k\alpha})$ with the
$w$s real and non-negative.  Then \expectv\ becomes
$$\eqalign{
\langle \Psi | H_{\rm int.} | \Psi \rangle =
{\sum_{k,k^\prime}} \,\,&4U (u_{k\uparrow} w_{k\uparrow}
u_{k^\prime\downarrow} w_{k^\prime\downarrow} )
 \cos \phi_{k\uparrow} \cos\phi_{k^\prime\downarrow}
\cr
&- 8V (u_{k\uparrow} w_{k\uparrow} \cos \phi_{k\uparrow}
 +u_{k\downarrow} w_{k\downarrow} \cos \phi_{k\downarrow})
(u_{k^\prime\uparrow} w_{k^\prime\uparrow} \cos \phi_{k^\prime\uparrow}
+u_{k^\prime\downarrow} w_{k^\prime\downarrow} \cos \phi_{k^\prime\downarrow})
\cr
&- 4V\, C({\bf k}-{\bf k^\prime})\,
\bigl( (u_{k\uparrow} w_{k\uparrow} \sin \phi_{k\uparrow}
u_{k^\prime\uparrow} w_{k^\prime\uparrow} \sin \phi_{k^\prime\uparrow})
\cr
&\qquad\qquad+ (u_{k\downarrow} w_{k\downarrow} \sin \phi_{k\downarrow}
u_{k^\prime\downarrow} w_{k^\prime\downarrow} \sin \phi_{k^\prime\downarrow})
  \bigr)~.
\cr}\eqn\realform$$

At this point, it is easy to identify three essentially different ways
whereby favorable correlation energy can emerge:

$\bullet$ From the first term, which favors $|\cos \phi | =1$ with opposite
signs for spin up and spin down -- this corresponds to antiferromagnetism
(this is the case analyzed in [\schrieffer]).

$\bullet$ From the second term, which favors $|\cos \phi | =1$ with the
same sign for spin up and spin down -- this corresponds to a charge density
wave.

$\bullet$ From the third term, which favors $|\sin \phi | = 1$ (that is,
$v$ pure imaginary relative to $u$).

The third possibility is in many ways the most interesting, and we shall
now discuss it in some detail
before returning to compare it with the other
two.

To continue the analysis we must minimize the full energy, including both
one-particle (hopping) and correlation parts.
For the one-particle part, we take the nearest-neighbor form
$\epsilon_k ~=~ -t(\cos k_x a + \cos k_y a)$.
For simplicity let us consider nearly half filling, so that
the chemical potential is small and the Fermi surface is near
the boundary of the magnetic zone.  The equations
for the
different spin components do not affect one another, so
we shall temporarily drop the spin index.  By standard methods
one then finds the solution of the minimization problem in the form
$$
\eqalign{
{u_k^2} &= {1\over 2}
\biggl(1- {\tilde\epsilon_k \over \sqrt {\tilde \epsilon_k^2 + \Delta_k^2}}
\biggr)
\cr
{w_k^2} &= {1\over 2}
  \biggl(1 + {\tilde \epsilon_k \over \sqrt {\tilde \epsilon_k^2 + \Delta_k^2}}
\biggr)\cr
}
\eqn\usandvs
$$
where $\tilde \epsilon_k \equiv {1\over2}(\epsilon_k - \epsilon_{k+Q})$ and
$$
\Delta_k \equiv  4V \sum_{k^\prime} u_{k^\prime} w_{k^\prime}
\sin \phi_k \sin \phi_{k^\prime}\, C({\bf k}-{\bf k^\prime})
\eqn\gapdef
$$
satisfies the gap equation
$$
\Delta_k ~=~ 2V \sum
{\Delta_{k^\prime} \over \sqrt {\epsilon_k^2 + \Delta_k^2}}
\sin \phi_k \sin \phi_{k^\prime}\,C({\bf k}-{\bf k^\prime}) ~.
\eqn\gapeqn
$$
and $C({\bf k}-{\bf k^\prime})$ can be written in the more revealing form
$$C({\bf k}-{\bf k^\prime})=
(\cos k_x a \cos k^\prime_x a + \cos k_y a \cos k^\prime_y a+
\sin k_xa \sin k^\prime_xa + \sin k_y a \sin k^\prime_y a )
\eqn\kernwrite$$

The energy will be minimized when $\Delta$ is non-negative;
together with the form of the gap equation this determines $\Delta_k$
to be of the form $\Delta_k = A |\sin k_x a | $ (or, what is
equivalent energetically, $\Delta_k = B |\sin k_y a |$)
which has a $p$-wave-like symmetry on the Fermi surface or
$\Delta_k = C |\cos k_x a |$ ($= C |\cos k_y a| $
on the diamond) which is $d$-wave-like.
The sign of $\sin \phi_k $ will adjust itself, changing
at the zeroes of $\Delta$ so
as to enforce these forms.  The gap equation is quite complicated
in general, but simplifies near half filling.  One has in this case
$$
1~=~
4V \int^\pi_0 {d\xi\over \sin \xi }
 \int d\tilde\epsilon{\sin^2 \xi\over\sqrt{\tilde \epsilon^2 + A^2 \sin^2 \xi}}
\eqn\Aform
$$
for the $A$ form and
$$
1~=~
8V \int^\pi_0 {d\xi\over \sin \xi }
 \int d\tilde\epsilon{\cos^2 \xi\over\sqrt{\tilde \epsilon^2 + C^2 \cos^2 \xi}}
 ~.
\eqn\Cform
$$
The ${1\over \sin \xi }$ factor, which reflects the van Hove singularity
in the density of states exactly at half filling, favors the $C$ form, as
does the factor 2 arising from the equality of the first two terms in the
final factor of \gapeqn. Away from half-filling, or in
the presence of a next-nearest neighbor hopping term, there will
not be a van Hove singularity at the Fermi surface.
However, an enhanced density of states will remain if these effects
are not too large. At half filling one has occupation up to the
edge of the diamond, where $\tilde \epsilon$ vanishes, and thus a solution
to the gap equation -- even apart from the van Hove enhancement -- for
arbitrarily weak coupling.

To discuss the situation below half filling, we must consider
how the condition
on electron density is implemented.  The operators
$$
\eqalign{
\gamma^\dagger_{k\alpha} &\equiv u_{k\alpha} c^\dagger_{k\alpha} +
v_{k\alpha} c^\dagger_{k+Q\,\alpha} \cr
\delta^\dagger_{k\alpha} &\equiv \bar v_{k\alpha}c^\dagger_{k\alpha} -
\bar u_{k\alpha} c^\dagger_{k+Q\,\alpha} ~,\cr
}
\eqn\quasiops
$$
analogous to the familiar Bogoliubov-Valatin
operators in superconductivity theory,
are a complete set of fermion creation and annihilation operators with
diagonal anticommutation relations.
$\gamma_{k\alpha}^\dagger$ and $\delta_{k\alpha}^\dagger$
create quasiparticles below and above the gap, respectively, or, in
semiconductor terminology, in the the valence
and conduction bands. At half-filling, all of the
$\gamma_{k\alpha}^\dagger$, or conduction band, states
are filled and the ground state \wavefnct\ can be written
$\Psi ~=~ \prod_k \gamma_{k\uparrow}^\dagger
\gamma_{k\downarrow}^\dagger |0>$, where ${\bf k}$ ranges
over all momenta in the magnetic zone. In general ${\bf k}$ ranges
only over momenta within the Fermi surface
(actually a curve, in two dimensions),
which is determined by the condition
$\sqrt {\tilde \epsilon_k^2 + \Delta_k^2}=\mu$.
The chemical potential, $\mu$, is chosen so that the
area enclosed by the Fermi surface is equal to
the desired density.
Below half-filling the $d \tilde \epsilon$ integral will be
cut off by $\mu$ instead of diverging as $C \rightarrow 0$.
Thus the gap equation will not have a
non-trivial solution at arbitrarily
weak coupling, but only starting at a finite value, dependent on the
deviation from half-filling, of the coupling.

Nothing in our considerations so far has correlated spin up and spin
down, so that spin singlet ($u_{k\uparrow}= u_{k\downarrow}$),
spin triplet ($u_{k\uparrow}=-u_{k\downarrow}$), or intermediate
possibilities
are equally favorable.  The degeneracy will be split if we include
into our Hamiltonian a nearest-neighbor spin-spin interaction
$J\vec s \cdot \vec s$
where of course
$\vec s (x)
\equiv c^\dagger_\alpha(x) \vec \sigma^\alpha_\beta c^\beta(x) $
is the electron spin operator.
When such a term is included in
the Hamiltonian, contributions sensitive to the pairing correlations
arise from the crossed channel.  A short calculation shows that
for $J>0$, antiferromagnetic coupling,
the spin singlet state is lowered in energy, and the triplet raised.
Of course the assumed effective coupling here might have either sign,
and it need not be at all the same as the effective $J$ used in a
different approximate description of the same material
({\it e.g}. in a $t-J$ model).   In any case the spin singlet order
will
be favored by its relative immunity from fluctuations, as we shall
presently discuss.

Now let us compare the energies of the competing orders. These
are quite complicated and depend on the interplay between
the single particle energies and the interaction terms
\expectv\ (with the $u_k$s and $v_k$s appropriate to these
orders substituted). Some qualitative features of the
phase diagram may be obtained
from the coefficients of the terms in the
purely repulsive Hamiltonian $H_{\rm int.}$.  We find that the
terms favoring the three types of states (antiferromagnetic,
charge density, ``C-type d-density'') are in the ratio
$-U: U - 8V: -4V\langle \cos^2 k_xa \rangle $. Here the last factor
arises from the non-trivial angular dependence of the d-density ordering
and the pairing interactions that give rise to it.  It will be
substantially larger
than ${1\over 2}$ for the $C$ form of d-density ordering, because the
density of states is largest where  the cosine is unity (indeed, to the
extent the van Hove singularity dominates, this factor is unity), though
its precise value is highly model- and doping-dependent.
When this factor is unity, the $C$ form of d-density ordering
is stable only at $U=4V$, according to this naive criterion.
However, one should also consider
that antiferromagnetic ordering, since it breaks a continuous symmetry, is
subject to severe fluctuation effects -- in principle, for example,
a two-dimensional system is rigorously forbidden to exhibit such ordering
at any non-zero temperature -- which can severely degrade the favorable
mean-field energy.
In a realistic domain of parameters for CuO$_2$ planes
Schrieffer {\it et al}.
[\schrieffer ] found that the magnetic moment density is renormalized by a
factor
.6 due to fluctuations, which is also approximately the factor
indicated by experiment.  One might expect the correlation energy to
be
renormalized roughly by the square of this factor.
The singlet d-density order, as we shall discuss further
below, breaks only a discrete symmetry, so it is safer in this regard.
Also the d-density order, because it contains nodes, has a more
favorable one-particle energy. The one-particle energy for
this state, in the presence of a gap of magnitude $\Delta$,
is $6.40 {\Delta^2}/t$ while the antiferromagnet or charge-density wave have
one-particle energy $10.65 {\Delta^2}/t$. (These numerical coefficients
were obtained by integrating the one particle energies up to
a chemical potential $\mu=2\Delta$ which avoids the van Hove singularity.
The results are similar for other choices of the chemical potential;
the robust point is that the d-density is substantially favored.)
Altogether, then, it seems that d-density order, especially in its
singlet form, is a serious candidate to describe real states of matter.

\chapter{Comments}

1. Since neither the Hamiltonian we analyzed nor the approximation
scheme we
employed is exact, it behooves us to identify what feature of
the proposed state might be expected to have precise validity.  As
in many cases in physics, a broken symmetry is the heart of the matter.
One finds in the spin-singlet d-density state
the non-vanishing expectation value
$$
\langle c^\dagger_\alpha (k+Q) c^\beta (k) \rangle ~=~
i \delta_\alpha^\beta  f(k)~,
\eqn\expval
$$
where $f$ is a real function that changes sign under a $\pi/2$
rotation.  The $i$ indicates that time-reversal symmetry $T$ is
violated, and the angular dependence indicates that symmetry under
$\pi/2$ rotation is violated.  Furthermore, the dependence on the
momentum offset $Q=(\pi/a, \pi/a)$ indicates that symmetry under
translation through
a single lattice spacing is also violated, since this too changes the
sign of the order parameter.  Let us swiftly add, however, that one
can combine any two of these broken symmetry operations to recover a valid
symmetry.  Thus, in particular, one does not expect to see any simple
direct
macroscopic manifestation of the symmetry breaking.

Indeed, it is not entirely easy to identify accessible signatures
for d-density order
even microscopically.  There are no extra coherent peaks in
elastic x-ray or neutron scattering, since the relevant single-particle
expectation values cancel upon integrating over angles.
Perhaps the most fundamental signature derives from
the existence
of
low energy particle-hole excitations at momenta connecting points on
the
Fermi surface repeated after translation through $Q$.  These should reveal
themselves
in inelastic neutron scattering, as an anomalous extended continuum.
Closely related structure would also be expected in photoemission.

Although the model we have analyzed above is two-dimensional, the
essential idea of the
ordering pattern \expval\ is not intrinsically two-dimensional, and
invites generalization to three-dimensional materials.

2.  We have already sketched the construction of the quasiparticles.
When one replaces the normal electron operators by these quasiparticles,
what emerges near half filling
can be described
to a first approximation using a semiconductor-like picture.  A gap has
opened up at the boundary of the magnetic zone, so that doping below
half-filling
it will produce an effective
concentration of hole-like charge carriers proportional
to the doping.  The Fermi surface will have its normal
area or volume, though if the nominal Fermi surface passes through the
magnetic zone boundary
it will be distorted from its nominal
band theory form. The critical behavior at the transition
will depend crucially on whether or not this occurs
[\millis,\sachdev].
Transport in the spin-triplet d-wave state will
be anomalous, due to the Nambu-Goldstone mode associated with breaking
of the spin rotation symmetry.  (More precisely, there will be
important fluctuations, but not necessarily a
simple Nambu-Goldstone mode, in
the appropriate channel.)   Transport in the
spin-singlet d-wave state will
also be anomalous, due to fluctuations associated
with transition to this state at a finite
coupling and doping.
The precise nature of these anomalies is under study.

Because it effectively halves the size of the Brillouin zone without
introducing any manifest antiferromagnetic or charge density ordering,
the d-density wave ordering appears to be an interesting candidate
to describe Mott insulators (or the related doped ``semiconductors'')
which exhibit neither
antiferromagnetic nor
charge-density ordering.

3.  Because the quasiparticles of
the d-density state are composed of d-wave mixtures of fundamental
electronic modes, the effective phonon-mediated interaction
between these particles will contain a substantial d-wave component.
Since it is less sensitive to Coulomb repulsion, effective
attraction in this
partial wave plausibly leads to a particularly robust instability
toward d-wave superconductivity.
(Of course, it is still subject to the strong fluctuations associated
with two-dimensionality for an isolated layer.)
The accumulation of density of states near the magnetic zone, which is
close to the Fermi surface for small doping, enhances this tendency.

4.  In the states we have considered above, the mean density of spin
up and spin down have been supposed equal.  Actually the $U$-term
energy (though not the single-particle energy) is minimized by a
completely polarized, ferromagnetic state.  Within the polarized state
one can still examine the possibility of d-density ordering, which
we find can indeed be energetically
favorable.  This provides a model for possible ferromagnetic Mott
insulator
behavior at $1\over 4$ filling.

\ack{We wish to thank R. Schrieffer and S. Sondhi for valuable
comments.}

\endpage

\refout

\end